# Being and Time in XR:

## Other-Presentness Beyond Co-Presence

**TOIDA Koichi**
MESON, Inc.
koichi.toida@meson.tokyo

**ABSTRACT** Research in XR (Extended Reality) has conventionally centred upon concepts such as Presence, Embodiment, Social Presence, and Co-presence. Within these traditions, bodily action, sensory contingencies, synchronous interaction, and possibilities for action have generally been regarded as constitutive conditions for the experience of "being there" and of being with others. XR environments, however, permit the partial separation of conditions that ordinarily co-vary in everyday experience. Bodily co-presence, temporal simultaneity, spatial configuration, and social interaction need not remain inseparable. This paper approaches this possibility as a problem of other-presentness. Other-presentness refers to the conditions under which another individual is experienced as existing "here and now". The contribution of this paper does not lie in arguing that asynchronous others can evoke social responses; such observations have already been addressed within parasocial interaction and social presence research. Rather, the novelty lies in theorising XR as a technological condition capable of separating and operationalising the constitutive elements of other-presentness as design variables. Reconsidering Bodyless Presence as a methodological precedent and drawing upon experimental findings from Immersive Video research, this paper formulates Bodyless Presentness as a condition in which another individual continues to be experienced as presently existing despite attenuated bodily co-presence and weakened real-time simultaneity.

**KEYWORDS** XR, Presence, Co-presence, Other-presentness, Bodyless Presentness, Phenomenology

### I. INTRODUCTION: The Problem of Other-Presentness

Research in XR has extensively examined Presence, Embodiment, Social Presence, and Co-presence (Slater and Wilbur, 1997; Slater, 2009; Skarbez et al., 2017). Within these traditions, bodily action, sensorimotor contingencies, synchronous interpersonal interaction, and possibilities for action have generally been understood as significant conditions supporting experiences of "being there" and of sharing experiences with others. XR, however, should not merely be understood as a technology for reproducing reality. XR environments permit the partial separation and recombination of bodily, spatial, temporal, and interpersonal conditions that ordinarily appear together in everyday experience. In ordinary circumstances, bodily co-presence, spatial proximity, gaze behaviour, and temporal simultaneity are typically encountered simultaneously. Within XR, by contrast, such conditions may be selectively weakened, preserved, or recombined.

For example, Immersive Video may induce the experience that another individual exists "here and now" despite being pre-recorded. Spatial Persona and volumetric telepresence permit remote individuals to appear within a shared space without physical co-location (Irlitti et al., 2023). Similarly, Substitutional Reality allows recorded events to be experienced not merely as representations of the past but as events unfolding within the present (Suzuki et al., 2012). A common characteristic of these examples is that the represented other need not be physically and temporally co-present in real time. Nevertheless, such others may continue to be experienced as presently existing. The central question may therefore be formulated as follows: must another individual be physically and temporally co-present in order to be experienced as existing "here and now"?

The present paper formulates this problem as other-presentness. In this paper, Presence refers primarily to the user's subjective sense of being situated within a mediated environment, whereas presentness refers to the experienced givenness of another individual as existing "here and now" within that environment. Other-presentness therefore designates not social engagement in general, but the specific experiential condition under which another individual is encountered as presently there. The argument developed here does not concern whether asynchronous others may evoke social responses. Such observations have already been demonstrated within parasocial interaction research (Horton

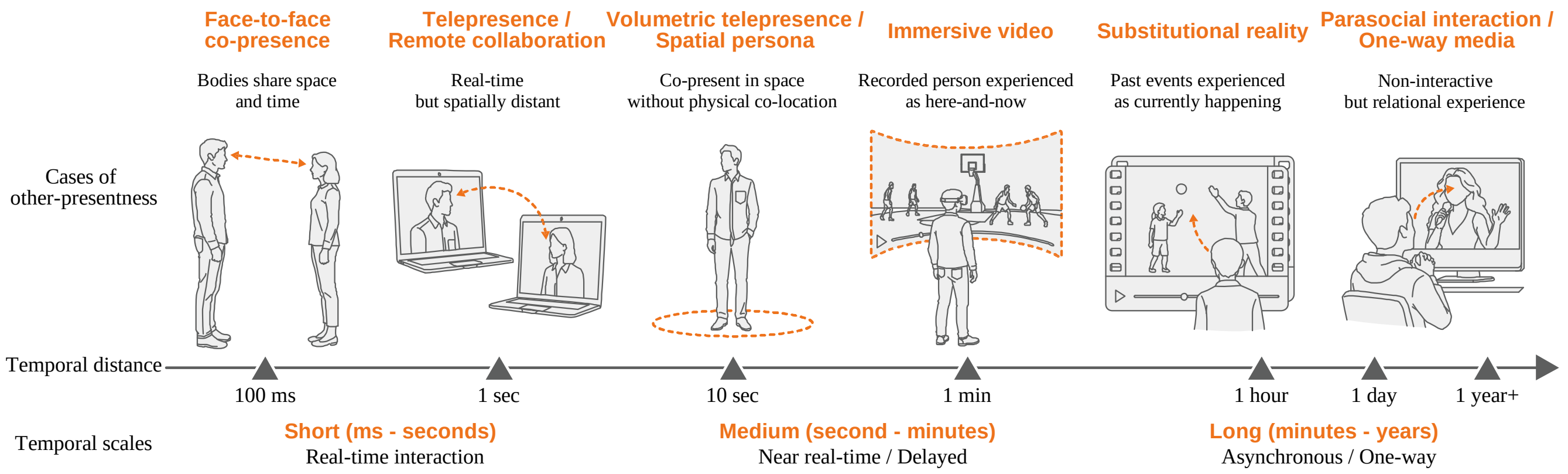

**FIGURE 1** Temporal scales of other-presentness across face-to-face, remote, recorded, and one-way mediated encounters in and around XR. The horizontal axis indicates conceptual temporal distance from reciprocal co-presence, not strict latency thresholds. Orange dashed elements indicate the dominant cues or structures through which other-presentness is stabilised in each case.

& Wohl, 1956; Tukachinsky & Stever, 2019). Rather, the concern lies elsewhere. XR permits the constitutive conditions of other-presentness to be separated and treated as spatial, temporal, and action-related variables. This possibility differs from earlier discussions of parasocial interaction or social presence. Parasocial interaction research demonstrated that mediated others may evoke forms of social engagement despite the absence of reciprocal interaction (Horton & Wohl, 1956; Tukachinsky & Stever, 2019). Social presence research similarly examined how media afford a sense of interpersonal connection (Biocca et al., 2003). However, these traditions largely addressed whether mediated others can become socially meaningful. XR introduces a different methodological possibility: conditions ordinarily coupled in everyday experience may be selectively preserved, weakened, or recombined. The question therefore shifts from whether mediated others are experienced socially to how constitutive conditions of other-presentness may be experimentally separated. It is this methodological characteristic that forms the principal object of inquiry.

## II. THEORETICAL BACKGROUND: Other-Presentness Across Temporal Scales

Social existence has not historically been examined within a single temporal scale. Different research traditions have instead investigated social experience across differing temporal structures. At shorter timescales, studies of Joint Attention and joint action have demonstrated that gaze behaviour, movement synchronisation, and response delay influence social interaction and co-presence (Frischen et al., 2007). Research concerning telepresence and remote collaboration has similarly shown that communication latency, spatial arrangement, and embodied representation influence experiences of Social Presence and Co-presence (Shin et al., 2022; Irlitti et al., 2023). At longer timescales, one-directional media and parasocial interaction research have demonstrated that enduring relationships with media figures may emerge despite the absence of reciprocal interaction (Horton & Wohl, 1956; Tukachinsky & Stever, 2019; Schramm et al., 2024). Consequently, interpersonal experience extends across temporal scales ranging from sub-second response delays to long-term mediated relationships.

XR complicates these distinctions by permitting multiple temporal structures to coexist within a single experiential environment. Recorded individuals, remote individuals, and asynchronously generated individuals may each appear within the same experiential space as presently existing others. The relevant issue therefore does not concern acceptable degrees of delay. Rather, it concerns the conditions under which another individual becomes experienced as presently existing across differing temporal scales (Fig.1).

An important empirical background is provided by prior Immersive Video research (Toida et al., 2026). Experimental work investigated whether filming distance altered Presence and whether such manipulations influenced psychological closeness towards performers. Idol performances were newly recorded under high-Presence and low-Presence conditions and subsequently experienced by actual fans. Manipulating filming distance significantly altered Spatial Presence, Self-Location, Social Presence, Possible Actions, and Observability. Higher-Presence conditions additionally produced larger increases in Inclusion of Other in the Self (IOS; Aron et al., 1992). Importantly, the study did not seek to measure fandom itself. Psychological closeness was instead operationalised specifically as perceived self-other overlap. Manipulating filming distance therefore altered not merely perceptual appearance but simultaneously affected self-location, social presence, action possibilities, and observability. These findings suggest that individuals represented within Immersive Video need not merely function as recorded objects. Rather, they may become experienced as others existing "here and now" through configurations of viewpoint position, distance, social presence, and action-related structures.

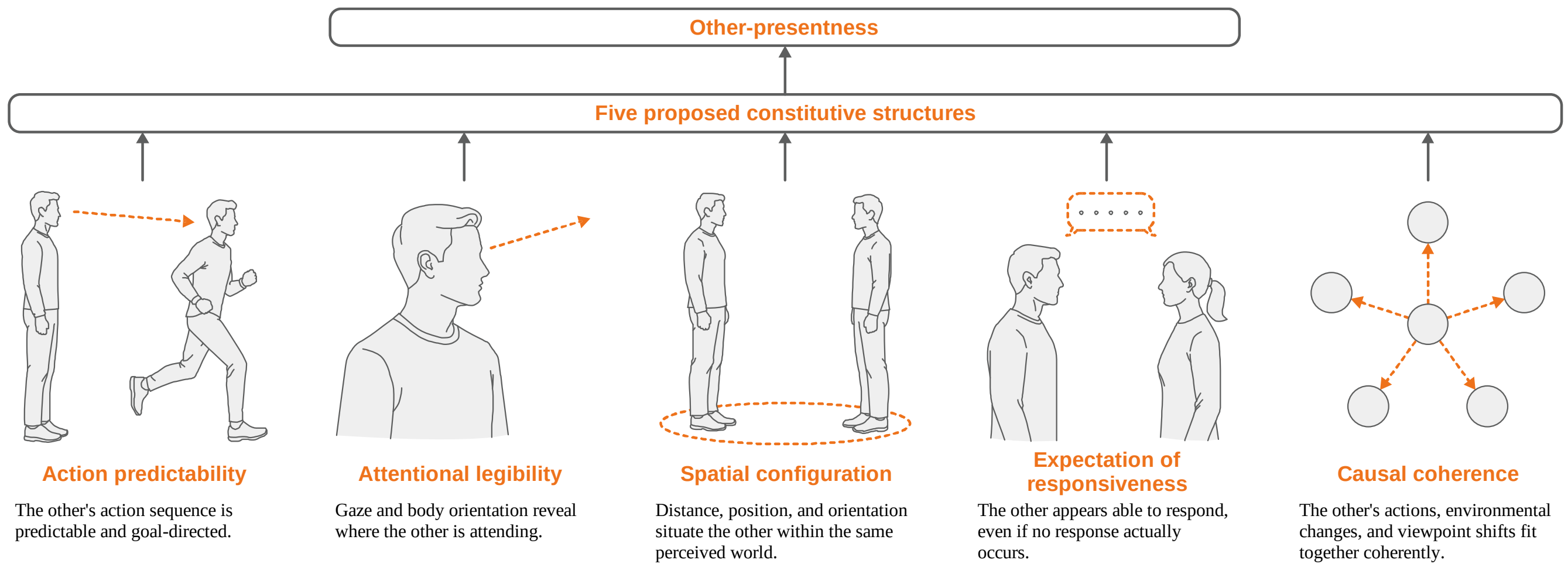

**FIGURE 2** Five proposed constitutive structures of other-presentness. The five structures jointly support the experience of another person as present, even when bodily co-location, real-time reciprocity, or direct interaction is partially weakened. Orange dashed elements indicate the salient cue emphasised in each structure.

## III. METHODOLOGICAL PRECEDENT: Bodyless Presence

Immersive Video has previously been positioned as a boundary case situated between interactive VR and conventional video (Toida, 2026a). Within this context, Presence was shown to emerge despite restricted locomotion, limited environmental interaction, and the absence of stable bodily representation. To account for this phenomenon, Bodyless Presence reconsidered relationships among agency, ownership, body schema, and self-location. Rather than proposing disappearance of body schema, the framework suggested a relative attenuation of body-schema availability accompanied by comparative dominance of self-location, described as a self-location-dominant state (Toida, 2026a).

The significance of Bodyless Presence did not lie in claiming that the body becomes unnecessary. Rather, its contribution lay in separating normally co-varying components of Presence and thereby exposing its constitutive conditions. If Bodyless Presence concerned conditions under which self-location persists despite attenuated body-schema availability, the present problem concerns conditions under which another individual remains experienced as presently existing despite weakened real-time co-presence.

## IV. TOWARDS BODYLESS PRESENTNESS: Engineering Other-Presentness

Substitutional Reality separates presentness from real-time simultaneity. Spatial Persona and volumetric telepresence separate physical co-location from shared space. Immersive Video separates recording from interpersonal proximity. These examples may therefore be understood not merely as technological demonstrations but as boundary conditions exposing constitutive elements of other-presentness. Existing work across social cognition, action understanding, interpersonal attention, and multisensory perception suggests that the experience of another as presently existing is unlikely to depend upon a single variable. Rather than emerging from bodily co-presence alone, interpersonal experience appears to arise through integration across multiple perceptual and predictive cues (Gallese, 2005; Frischen et al., 2007; Shams & Beierholm, 2010). XR becomes theoretically important because these ordinarily co-varying structures may be experimentally decoupled. The five structures proposed below are not intended as an exhaustive or mutually orthogonal taxonomy. Rather, they are proposed as analytically separable candidate structures that repeatedly appear across research on action understanding, attention, spatial social interaction, mediated reciprocity, and causal inference. Their value lies less in claiming completeness than in providing a set of experimentally manipulable dimensions through which XR-mediated other-presentness can be described, compared, and eventually tested.

First, action predictability concerns the extent to which observed bodily movements are organised as coherent, goal-directed actions rather than as merely visual motion. Such action legibility may be supported by mirror-neuron, embodied-simulation, and sensorimotor-integration accounts, which suggest that observed actions can be mapped onto the observer’s own motor or bodily representations (Rizzolatti & Craighero, 2004; Gallese, 2005; Shimada, 2022). Second, attentional legibility concerns the extent to which gaze behaviour and bodily orientation make the other's attentional direction readable. Such cues organise social meaning and guide observers' attention (Frischen et al., 2007). Third, spatial configuration concerns the extent to which distance, orientation, and positioning make another individual appear to belong within the same experiential world (Shin et al., 2022; Irlitti et al., 2023). Fourth, expectation of responsiveness concerns the extent to which the other appears capable of responding,

| | Five proposed constitutive structures: Action predictability | Attentional legibility | Spatial configuration | Expectation of responsiveness | Causal coherence | Typical constraint separated |
|---|---|---|---|---|---|---|
| Face-to-face co-presence | ◎ | ◎ | ◎ | ◎ | ◎ | None (baseline) |
| Telepresence / Remote collaboration | ◎ | ◎ | ○ | ◎ | ○ | Spatial co-location |
| Volumetric telepresence / Spatial persona | ○ | ○ | ◎ | ○ | ○ | Physical co-presence |
| Immersive video | ○ | ○ | ○ | △ | ○ | Real-time simultaneity (recorded) |
| Substitutional reality | ○ | ○ | ○ | △ | △ | Real-time simultaneity (currentness) |
| Parasocial interaction / One-way media | △ | △ | ○ | ○ | △ | Reciprocity (interactivity) |

◎: Strongly supported ○: moderately supported △: weakly supported ×: Not supported / Absent even when actual interaction does not occur (Horton & Wohl, 1956). Fifth, causal coherence concerns the extent to which the other's actions, environmental changes, and events are inferred as originating from common causes (Körding et al., 2007; Shams & Beierholm, 2010). Together, these five structures constitute a proposed framework of other-presentness (Fig. 2).

**FIGURE 3** Selective preservation and attenuation of other-presentness structures across XR-mediated others. The ratings indicate qualitative, theory-driven judgements based on the typical configuration of each case, rather than empirical measurements. They are intended to illustrate relative preservation or attenuation of the proposed structures and should be treated as hypotheses for future empirical validation.

Importantly, these structures need not contribute equally. Bodily co-presence in everyday environments may function as a compression mechanism through which multiple cues become simultaneously available. XR environments, however, permit selective preservation of some cues while attenuating others. A recorded performer may preserve spatial configuration and attentional legibility while lacking reciprocity. A remote collaborator may preserve responsiveness while weakening bodily realism. Synthetic agents may maintain causal coherence without biological authenticity. Consequently, different forms of mediated others may occupy distinct regions within a shared space of other-presentness. In this context, bodyless does not mean that the observer lacks a body, nor that the represented other is visually disembodied. Rather, it indicates that bodily co-presence between self and other is attenuated as a constitutive condition of interpersonal presentness. Bodyless Presentness may therefore be defined as a condition in which another individual continues to be experienced as existing "here and now" despite attenuated bodily co-presence and weakened real-time simultaneity.

## V. DISCUSSION AND IMPLICATIONS: Beyond Co-Presence

Other-presentness cannot be straightforwardly reduced to bodily co-presence or temporal simultaneity. Although bodily co-presence and shared time remain powerful conditions supporting interpersonal experience, their attenuation does not necessarily eliminate the experience of another as presently existing. This perspective first repositions telepresence research. Traditional telepresence studies have frequently prioritised communication fidelity, latency reduction, avatar realism, and collaborative performance. From the perspective of other-presentness, however, the central issue shifts towards identifying conditions under which another individual becomes experienced as existing "here and now". This shift is not limited to conventional telepresence. It also reframes emerging discussions concerning artificial or synthetic social agents. Work on human responses to media and computers has long suggested that users may respond socially to mediated entities even when they know that these entities are technologically produced (Reeves & Nass, 1996). From the perspective of other-presentness, the central issue is therefore not whether an entity is biologically human, nor whether it is perfectly realistic, but which conditions sustain the experience that another is presently there. A recorded performer, a remote collaborator, and a synthetic agent may differ ontologically and technically, yet each may be analysed in terms of action predictability, attentional legibility, spatial configuration, responsiveness, and causal coherence.

Second, Bodyless Presentness provides a shared analytical framework for comparing heterogeneous forms of mediated others without reducing them to either technological format, biological authenticity, or real-time simultaneity (Fig. 3). This framework is also empirically testable. Each proposed structure can be selectively manipulated while measuring subjective other-presentness, social presence, perceived co-presence, behavioural orientation towards the other, and physiological or attentional responses. For example, immersive video could vary gaze direction, interpersonal distance, action continuity, implied responsiveness, or causal synchrony between the other's actions and environmental events. If other-presentness remains high despite weakened real-time reciprocity, but decreases when causal coherence or spatial configuration is disrupted, the framework would receive partial support. Conversely, if these manipulations fail to dissociate other-presentness from conventional co-presence or social presence measures, the framework would require revision.

Third, this framework extends Presence research itself. Just as Bodyless Presence reconsidered Presence beyond bodily ownership and agency, Bodyless Presentness reconsiders interpersonal presence beyond bodily co-location and temporal simultaneity. The question is not merely whether the user feels present within a mediated environment, but how another becomes present within that environment. This shift moves from the self-location of the user towards the presentness of the other. Yet if another can be experienced as present “now” without simultaneity or reciprocity, a more fundamental question emerges: what makes this “now” a form of first-personal experience rather than merely a property of shared time—a question explored in *Present After Presence* (Toida, 2026b).

Heidegger (1927/1962) described *Mitsein* not as simple physical co-location but as a structure through which existence is always already shared with others within a common world. XR environments expose this structure under altered boundary conditions. In this sense, XR does not merely provide new media forms. It alters the conditions under which coexistence itself becomes observable as a structured phenomenon. When bodily co-presence and temporal simultaneity become attenuated while another nevertheless continues to appear as presently existing, the question concerns what continues to sustain the experience of being-with. XR should therefore not be understood merely as a technology for reproducing others. XR permits the constitutive conditions through which another individual becomes experienced as existing "here and now" to be partially separated and reconsidered. In this sense, XR functions not only as a medium of Presence but also as a medium through which conditions of Presentness become newly visible. Bodyless Presentness can therefore be understood as an XR-specific way of making the conditions of being-with experimentally visible. In ordinary face-to-face situations, being-with is supported by the dense co-occurrence of bodily presence, shared temporality, spatial proximity, mutual orientation, and potential responsiveness. XR loosens this density. It allows some of these conditions to be preserved while others are weakened or removed. The significance of XR for the question of Mitsein is therefore not that it replaces philosophical analysis, but that it creates technical situations in which the normally tacit conditions of being-with can be separated, recombined, and observed.

## ACKNOWLEDGMENT

The author acknowledges all past and present collaborators whose discussions and intellectual exchanges have shaped the broader perspective from which this work was developed. The ideas presented here did not emerge in isolation, but always already in a world shared with others.